\documentclass[prb,floatfix,twocolumn,showpacs,amsmath,amssymb]{revtex4}

\usepackage{graphicx}
\usepackage{dcolumn}
\usepackage{bm}

\begin{document}
\title{
Zero modes of tight binding electrons on the honeycomb lattice
}
\author{Yasumasa Hasegawa$^{1}$, Rikio Konno$^{2}$, Hiroki Nakano$^{1}$,
and Mahito Kohmoto$^{3}$}
\affiliation{
$^{1}$
Graduate School of Material Science,
University of Hyogo, Hyogo 678-1297, Japan\\
$^{2}$Kinki University Technical College,
2800 Arima-cho, Kumano-shi, Mie 519-4395,
Japan \\
$^{3}$Institute for Solid State Physics, University of Tokyo,
 Chiba, 277-8581, Japan
}

\date{\today}


\begin{abstract}
Tight binding electrons on the honeycomb lattice are 
studied where nearest neighbor hoppings in the three directions are $t_a,t_b$ and $t_c$, 
respectively. For the isotropic case, namely for $t_a=t_b=t_c$, two zero modes 
exist where the energy dispersions at the vanishing points are linear in momentum $k$.
Positions of zero modes move in the momentum space as $t_a,t_b$ and $t_c$ are varied.
It is shown that  zero modes exist if  $\left|\left|\frac{t_b}{t_a}\right| -1\right| \leq
\left|\frac{t_c}{t_a}\right| \leq \left|\left|\frac{t_b}{t_a}\right|+1\right|$. The density of states near a zero mode is proportional to $|E|$ but it is propotional to $\sqrt{|E|}$ at the boundary of this condition

\hspace{2cm}
\end{abstract}

\maketitle


The integer quantum Hall effect has been observed in graphene\cite{Novoselov2005, Zhang2005} when
the carriers are changed by the gate voltage. 
The quantization of the Hall effect is observed as $\sigma_{xy}=2 n\frac{e^2}{h}$ with $n= \pm 1,
\pm 3 \cdots$, where the factor $2$ comes from the spin degrees of freedom.
These quantum numbers are unusual, since in a usual case $n=0, \pm1, \pm2, \cdots$.
This unusual quantum Hall effect was discussed in terms of relativistic Dirac theory \cite{Gusynin2005}. 
However it is  more natural to be explained by
the realization of the quantum Hall effect in periodic systems\cite{TKNN1982}
in the presence of zero modes\cite{Hasegawa2006,Peres2006}.
 We will call zero modes instead of massless Dirac excitations in this paper because
we do not consider relativistic particles.
The energy spectrum and the density of states of the honeycomb lattice near half filling and
in zero or small magnetic field are similar to these in the square lattice   near half filling
in a very strong magnetic field about half flux quantum per each unit cell\cite{Hasegawa2006}.

At zero carrier concentration (i.e. half-filled electrons),
the resistivity $\rho_{xx}$ is close to the quantum value $h/(4e^2)=6.45 k \Omega$   
independent of temperature\cite{Novoselov2005},
which has been also attributed to the zero modes\cite{Novoselov2005, Zhang2005,Ziegler1998}.

 The existence of zero modes has also been proposed for the quasi-two-dimensional 
organic conductor
$\alpha$-(BEDT-TTF)$_2$I$_3$. The conductivity  under pressure is almost constant in a wide range 
of temperature\cite{Tajima2002}. Pertinent numerical computations performed by
 Kobayashi et al.\cite{Kobayashi2004} found that, for certain range of parameters,
the Fermi surfaces
become points and the density of states is proportional to
energy at 3/4 filling of electrons.
The existence of zero modes
was also  confirmed by the band structure calculation\cite{Ishibashi2006,Kino2006}. 
The unit cell for the model of $\alpha$-(BEDT-TTF)$_2$I$_3$ has four non-equivalent sites.
Katayama et al.\cite{Katayama2006} studied simpler model 
with two sites in the unit cell and they obtained a condition for zero modes.

\begin{figure}[tbh]
\includegraphics[width=0.46\textwidth]{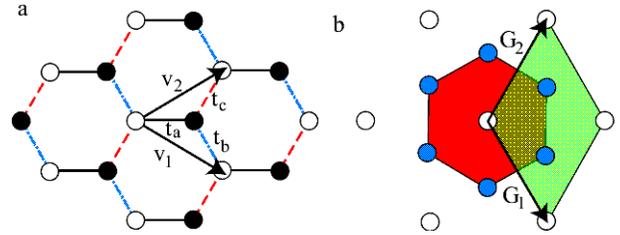}
\caption{
(color online) a. honeycomb lattice. 
Unit vectors are $\mathbf{v}_1=(\frac{3a}{2}, -\frac{\sqrt{3}a}{2})$
and $\mathbf{v}_2=(\frac{3a}{2}, \frac{\sqrt{3}a}{2}$).
Three  nearest neighbor  hoppings are $t_a$, $t_b$ and $t_c$ b. 
The red hexagon is a Brillouin zone for the honeycomb lattice. 
The reciprocal lattice vectors are 
$\mathbf{G}_1=(\frac{2\pi}{3a}, -\frac{2\pi\sqrt{3}}{3a})$
and $\mathbf{G}_2=(\frac{2\pi}{3a}, \frac{2\pi\sqrt{3}}{3a})$.
White circles are $\Gamma$ points. Brillouin zone can also be taken by the green diamond.
}
\label{fig1ab}
\end{figure}

In this Letter we study a tight binding model on the honeycomb lattice 
and obtain the condition of $t_a,t_b$ and $t_c$ for the existence of zero modes.

Unit cell of the honeycomb lattice contains two sublattices as shown 
in Fig. \ref{fig1ab}a. The Bravais lattice is a triangular lattice with 
\begin{align}
 \mathbf{v}_1 &= (\frac{3a}{2}, -\frac{\sqrt{3}a}{2} ), \\
 \mathbf{v}_2 &= (\frac{3a}{2}, \frac{\sqrt{3}a}{2} ), 
\end{align}
where $a$ is a distance between nearest sites.
We consider only nearest neighbor hoppings. There are three 
nearest neighbors for each site, $t_a$, $t_b$ and $t_c$ as shown  in Fig.\ref{fig1ab}.
We study the generalized honeycomb lattice model where $t_a$, $t_b$ and $t_c$ are not necessarily equal.
Under uniaxial pressure, $t_a$, $t_b$ and $t_c$ have different values for each other.
For example, $t_a > t_b = t_c$ is expected, if the uniaxial pressure along the $x$ direction is applied.
The  Hamiltonian for the generalized honeycomb lattice is given by
\begin{align}
  \mathcal{H}= \sum_{\mathbf{r_m}} \Big[ 
  &-t_a (a^{\dagger}_{\mathbf{r}_m} b_{\mathbf{r}_m} + h.c.) \nonumber \\
  &-t_b (a^{\dagger}_{\mathbf{r}_m+\mathbf{v}_1} b_{\mathbf{r}_m} + h.c.) \nonumber \\
  &-t_c (a^{\dagger}_{\mathbf{r}_m+\mathbf{v}_2} b_{\mathbf{r}_m} + h.c.) \Big] .
\end{align}
Using the Fourier transform 
\begin{align}
  a_{\mathbf{r}_m} &= \sum_{\mathbf{k}} e^{-i \mathbf{k} \cdot \mathbf{r}_m} a_{\mathbf{k}}, \\
  b_{\mathbf{r}_m} &= \sum_{\mathbf{k}} e^{-i \mathbf{k} \cdot (\mathbf{r}_m + \mathbf{x})} b_{\mathbf{k}},
\end{align}
where $\mathbf{x} = (a,0)$, 
we obtain
\begin{align}
  \mathcal{H} &= \sum_{\mathbf{k}}\Bigg[ \Bigg( -t_a \exp({-ik_x}) 
       - t_b \exp \left (i \left(\frac{1}{2}k_x-\frac{\sqrt{3}}{2}k_y\right)\right)  \nonumber \\
 &  -t_c \exp \left(i \left(\frac{1}{2}k_x + \frac{\sqrt{3}}{2}k_y \right) \right) \Bigg) 
  a^{\dagger}_{\mathbf{k}} b_{\mathbf{k}} + h.c.\Bigg] .
\end{align}
The energy is given by
\begin{align}
 \epsilon_{\mathbf{k}}^2&=  
 t_a^2+t_b^2+t_c^2 \nonumber \\
 &+ 2t_a t_b \cos \left(\frac{3}{2}k_x - \frac{\sqrt{3}}{2} k_y \right)
 + 2 t_a t_c \cos\left( \frac{3}{2}k_x + \frac{\sqrt{3}}{2} k_y \right)
 \nonumber \\
 &+ 2 t_b t_c \cos \left( \sqrt{3} k_y \right).
\end{align}
If we  perform a translation in the momentum space
\begin{equation}
(k_x, k_y) \rightarrow (k_x+\frac{2}{3}\pi,k_y),
\end{equation}
 and a replacement $t_a \rightarrow -t_a$ simultaneously,
we get the same $\epsilon_{\mathbf{k}}$.
Therefore we can take $t_a \geq 0$ without loss of generality. 
In a similar way one can take $t_b \geq 0$
and $t_c \geq 0$ without loss of generality by taking a translation in the momentum space,
\begin{equation}
(k_x, k_y) \rightarrow (k_x+\frac{1}{3}\pi,k_y \pm \frac{\sqrt{3}}{3}\pi).
\end{equation}
The reciprocal lattice vectors are
\begin{align}
  \mathbf{G}_1 &= (\frac{2\pi}{3a}, -\frac{2\pi\sqrt{3}}{3a}), \\
  \mathbf{G}_2 &= (\frac{2\pi}{3a},  \frac{2\pi\sqrt{3}}{3a}),
\end{align}
as shown in Fig.~\ref{fig1ab}b.
Let us write 
\begin{equation}
 \mathbf{k} = k_1 \mathbf{G}_1 + k_2 \mathbf{G}_2,
\end{equation}
where
\begin{align}
 k_1 &= \frac{3}{4\pi} k_x  - \frac{3}{4\sqrt{3}\pi} k_y, \\
 k_2 &= \frac{3}{4\pi} k_x + \frac{3}{4\sqrt{3}\pi} k_y.
\end{align}
The energy is 
\begin{align}
 \epsilon_{\mathbf{k}}^2 &= t_a^2+t_b^2+t_c^2 + 2t_a t_b \cos(2\pi k_1)
\nonumber \\
&+ 2t_a t_c \cos(2\pi k_2) + 2t_b t_c \cos (2 \pi (-k_1+k_2)). 
\end{align}
\begin{figure}[tb]
\includegraphics[width=0.4\textwidth]{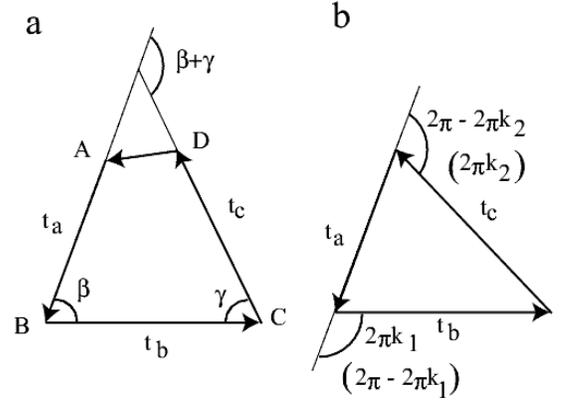}
\caption{Graphical explanation for appearance of zero modes. (a) If
$t_a$, $t_b$ and $t_c$ do not form a triangle, there are no
zero modes and gaps at $E=0$ are open. 
(b)  Zero modes exist when $t_a$, $t_b$ and $t_c$ form a triangle. Angles $k_1$ and $k_2$
are shown.}
\label{fig3}
\end{figure}

\begin{figure}[tbh]
\includegraphics[width=0.3\textwidth]{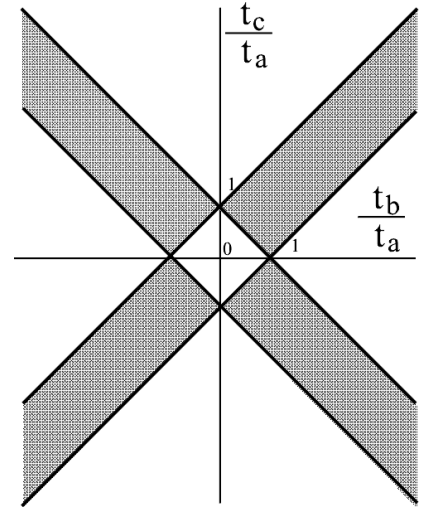}
\caption{Zero modes exist in the filled region. }
\label{fig4}
\end{figure}

\begin{figure}[tbh]
\includegraphics[width=0.46\textwidth]{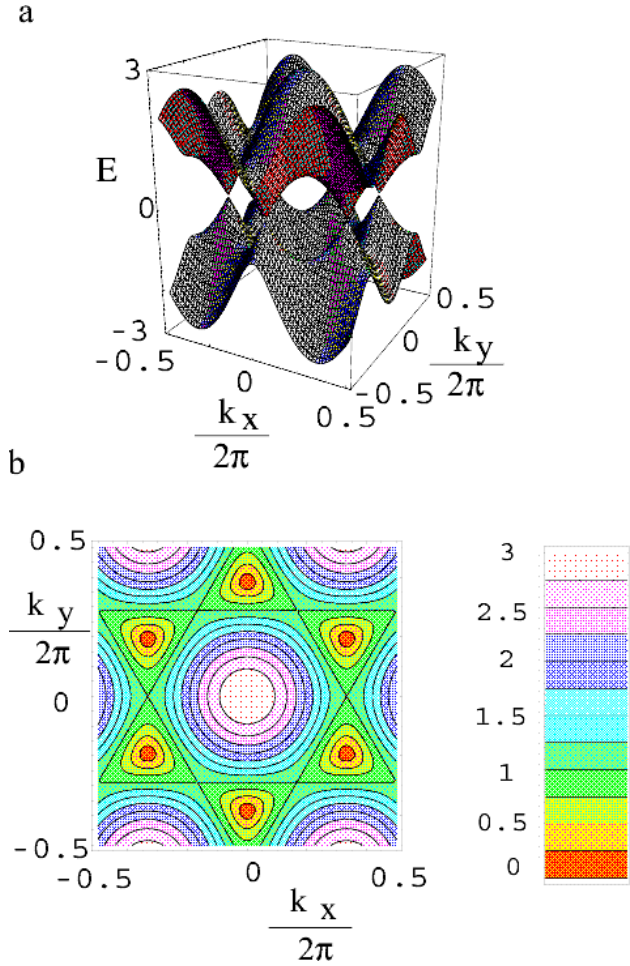}
\caption{(color online) 
Energy dispersion for the isotropic case  ($t_a=t_b=t_c=1$).
(a) 3D plot and  (b) contour plot.}
\label{fighoney3d1110}
\end{figure}
\begin{figure}[tbh]
\includegraphics[width=0.46\textwidth]{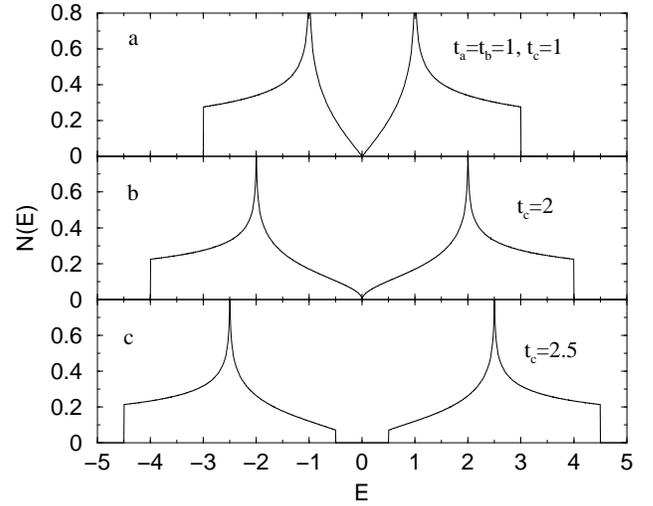}
\caption{Density of states of the electrons on a generalized honeycomb lattice.
}
\label{figdos}
\end{figure}
\begin{figure}[tbh]
\includegraphics[width=0.46\textwidth]{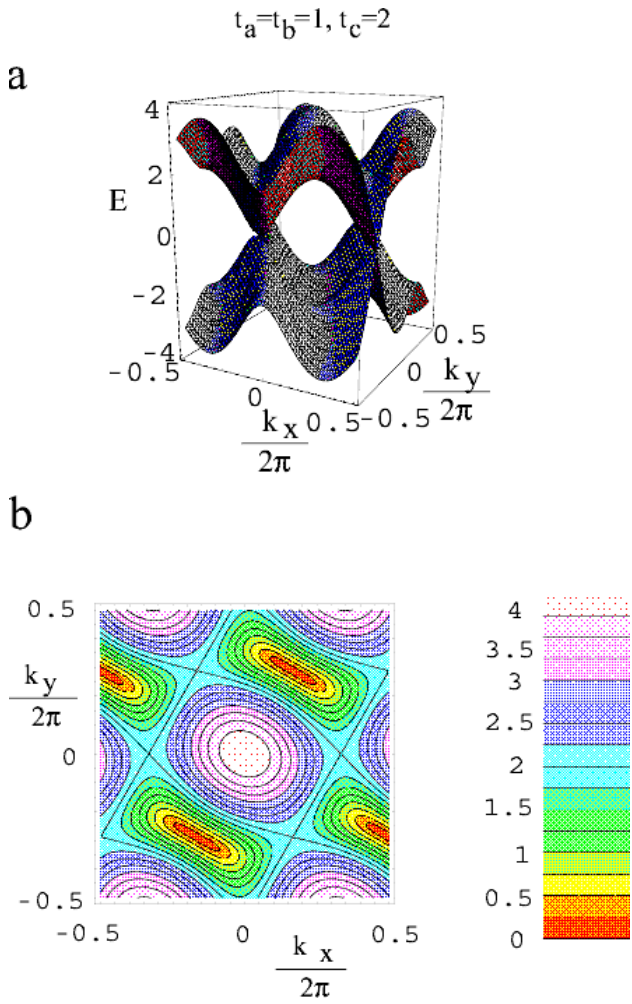}
\caption{(color online) 
3D plot (a) and the contour plot (b) of the energy of the generalized honeycomb 
lattice with $t_a=t_b=1$, $t_c=2$.}
\label{fighoney3d1120}
\end{figure}
The minimum of $|\epsilon_{\mathbf{k}}|$ is obtained as follows.
Consider the quadrangle ABCD in Fig.~\ref{fig3}a. Then we have
\begin{align}
  |\overrightarrow{DA}|^2 &= |\overrightarrow{AB} 
+  \overrightarrow{BC}      + \overrightarrow{CD}|^2 \nonumber \\
 &= t_a^2 +t_b^2 +t_c^2  \nonumber \\
 & -2 t_a t_b \cos \beta -2 t_b t_c \cos \gamma -2 t_a t_c \cos (\pi-\beta -\gamma)  \geq 0.
\end{align}
Put
\begin{align}
 \beta &= \pi-2 \pi k_1 \\
 \gamma &= \pi + 2\pi (k_1-k_2),
\end{align}
then 
\begin{align}
& t_a^2+t_b^2+t_c^2 +2t_a t_b \cos (2 \pi k_1) \nonumber \\
+& 2 t_a t_c \cos (2\pi k_2) + 2 t_b t_c \cos (2\pi (k_1 -k_2)) \geq 0
\end{align}
The equality is satisfied when $t_a$, $t_b$ and $t_c$ form a triangle which can be seen in Fig.~\ref{fig3}b.,
i.e.,
\begin{eqnarray}
  \cos(2\pi k_1)      &= \frac{t_c^2-t_a^2-t_b^2}{2 t_a t_b} \\
  \cos(2\pi k_2)      &= \frac{t_b^2-t_a^2-t_c^2}{2 t_a t_c} \\
  \cos(2\pi (k_1-k_2))&= \frac{t_a^2-t_b^2-t_c^2}{2 t_b t_c}
\end{eqnarray}
The triangle can be formed when

\begin{equation}
 \left|\frac{|t_b|}{|t_a|} -1 \right| \leq \frac{|t_c|}{|t_a|} \leq  \left|\frac{|t_b|}{|t_a|} +1 \right|,
\label{condition}
\end{equation}
is satisfied.
See Fig.~\ref{fig4}.

In the isotropic case where $t_a=t_b=t_c$,
zero modes are at
$(k_1,k_2)=\pm (\frac{1}{3}, \frac{2}{3})$,
 $(k_1,k_2)=\pm (\frac{2}{3},\frac{1}{3})$,  and $(k_1,k_2)=\pm (-\frac{1}{3},\frac{1}{3})$
i.e. the corners of the first Brillouin zone,
$(k_x,k_y)=\pm (\frac{2\pi}{3}, \frac{2\sqrt{3}\pi}{9})$,  
$(k_x,k_y)=\pm (\frac{2\pi}{3}, -\frac{2\sqrt{3}\pi}{9})$
and $(k_x,k_y)=\pm (0, \frac{4\sqrt{3}\pi}{9})$. See Fig.~\ref{fighoney3d1110}.
The density of states is plotted in Fig.~\ref{figdos}a.

If the parameters are in the boundary as seen in Fig.~\ref{fig4}, two zero modes 
merge into a confluent point.
For example, $\epsilon_k=0$ at confluent point $(k_{1}^*,k_2^*)=(0,1/2)$ for $t_a=t_b=1$, $t_c=2$
(Fig.~\ref{fighoney3d1120}).
Near this point $\epsilon_{\mathbf{k}}$ is written as
\begin{equation}
 \epsilon_{\mathbf{k}} \propto \pm \sqrt{c_1 (k_1-k_1^*)^4 + c_2 (k_2-k_2^*)^2}
\end{equation}
where $c_1$ and $c_2$ are constants.
In this case the density of states near $E=0$ becomes 
\begin{equation}
N(E) \propto \sqrt{|E|}.
\end{equation}
 See Fig.~\ref{figdos} b),
while $N(E) \propto |E|$ in the case of two zero modes (Fig.~\ref{figdos}a).
When the inequality Eq.(\ref{condition}) is not satisfied,
a finite gap opens at $E=0$ as shown in Fig.~\ref{figdos}c.

In conclusion, we have studied the energy of 
tight binding electrons in the generalized honeycomb lattice and found the
condition for the existence of zero modes.
The zero modes exist at the corners of the hexagonal first Brillouin zone for the usual honeycomb lattice.
Two zero modes moved to become a confluent point at the critical values of parameters $t_a$, $t_b$ and $t_c$,
where $t_a$, $t_b$ and $t_c$ stop to form a triangle.


This work is supported by a Grant-in-Aid 
for the Promotion of Science and Scientific Research on
Priority Areas (Grant No. 16038223) from the Ministry of
Education, Culture, Sports, Science and Technology.

%
%
\newpage 

\end{document}